\documentclass[twocolumn]{aa}
\usepackage{graphicx}
\usepackage{aalongtable}
\usepackage[figuresright]{rotating}
\usepackage{txfonts}
\def\grau{$^\circ$}

\def\nodata{...}
\begin{document}
   \title{Nebular abundances of southern symbiotic stars\thanks{Based on observations made at Observat\'orio do Pico dos Dias / LNA (Brazil) and European Southern Observatory (Chile)} }


   \author{G. J. M. Luna
          \inst{1}
          \and
          R.D.D. Costa\inst{1}
                    }

   \offprints{G. J. M. Luna}

   \institute{Instituto de Astronomia, Geofisica e Ci\^encias Atmosf\'ericas  (IAG), Universidade de S\~ao Paulo, S\~ao Paulo, Brazil\\
              \email{gjmluna@astro.iag.usp.br, roberto@astro.iag.usp.br}
             }

   \date{ }

   \abstract{
   We have calculated relative elemental abundances for a sample of 43 symbiotic stars. 
Helium abundances and the relative elemental abundances N/O, Ne/O, Ar/O were derived from 
new spectra collected in the optical range through low dispersion spectroscopy.
The He ionic abundances were derived taking into account self-absorption effects in Balmer lines. 
We found that the symbiotic stars in the galactic bulge have heavy element abundances 
showing the same wide distribution as other bulge objects. In the galactic disk, the 
symbiotic stars follow the abundance gradient as derived from different kinds of objects.
 
   }

  \maketitle

%

\section{Introduction}

   Symbiotic stars are binary systems with large periods and strong interaction. There is an agreement in the fact that they consist of (at least) three components: a giant star, a hot source like a white dwarf, a main sequence star or even a neutron star (GX1+4), and a nebula ejected by the red giant, as was shown by Nussbaumer et al. (1988). The nebula can be ionized by the UV radiation from the hot source and in some eruptive symbiotic stars also by the region where the winds from the hot and cold sources collide (\cite{willson}).
As the emission lines are very strong, the symbiotic stars are easily observable at large distances and are useful tools to test some aspects of the chemical composition of the low and intermediate mass population of the disk and galactic bulge as well as the evolution of double stars.

Some studies on nebular abundances in symbiotic stars have been performed in the optical region (see \cite{costa}, Pereira et al. 2002, \cite{moreno}) but with a small number of objects. Medina Tanco \& Steiner (1995) have made spectral classification of a sample of symbiotic stars toward the galactic bulge, but they did not derive chemical abundances. In the UV region, CNO abundances were derived for expressive samples of symbiotics using IUE data (e.g. Nussbaumer et al. 1988, Schmid \& Nussbaumer 1993).

The analysis of chemical abundances is required to investigate the surface enrichment of the red giant photosphere, whose stellar wind reflects the modifications introduced by dredge up processes along the stellar evolution. In this case, the investigation can be performed through techniques developed to study emission nebulae, which allow the determination of chemical abundances of elements such as helium, nitrogen, oxygen, neon and argon. Helium abundances must be derived with some caution, because the metastability of the 2$^{3}$S level causes radiative transfer effects and induces collisional excitation which can affect the final result. A second problem arises from the use of the Balmer decrement for reddening correction. The observed values suggest self-absorption effects in some systems which must be taken into account.

In this work we report the derivation of relative elemental abundances for 43 southern symbiotic stars. In sections 2 to 4 we discuss the observation and reduction techniques, in sections 5 to 7 the methods of analysis are described, and in section 8 the results are discussed.

\section{The sample}

Our sample was selected from the \cite{belczynski} catalog of symbiotic stars, and one object (SS73 71) was added from \cite{pereira}. As a selection criteria we chose all the symbiotic stars toward the galactic bulge, which we have roughly defined as the region between 20\grau $\leq$ \textit{l} $\leq$ 20\grau and 20\grau $\leq \textit{b} \leq$ 20\grau. With these criteria our sample has 90 objects. We cannot ensure that all of them belong to the galactic bulge, because the lack of good distances for the sample, but clearly all of them belong to the intermediate age population of the disk/bulge regions. Additionally, we have observed some other objects from the \cite{belczynski} catalog, that are out of our bulge definition.

\section{Observations}

  Spectroscopic observations were performed in two runs at the National Laboratory for Astrophysics (Bras\'opolis, MG, Brazil) from 07-16/Jun/2002, and from 23-26/Jun/2003, using a Boller \& Chivens Cassegrain spectrograph attached to the 1.60m telescope with a dispersion of 4.4 \AA/pixel. Some observations were made at ESO using the 1.52m telescope in La Silla, Chile (from 08-13/Oct/2002) with a Boller \& Chivens Cassegrain spectrograph with a dispersion of 2.2 \AA/pixel. Spectra cover the range 3800-7400 \AA. The log of the observations can be seen in the Table 1.
Each object was observed at least twice in the corresponding observational run, one of them with a short exposure time to get the fluxes of H$\alpha$, H$\beta$, H$\gamma$ and H$\delta$ to derive the reddening correction, and the other with a longer exposure time, saturating the Balmer lines and getting the weaker line fluxes. All the observations were performed in weather conditions compatible with flux calibration, and with an average seeing of 2 arcsecs for LNA and 1 arcsec for ESO.
Flux calibration for each object was secured through observations of standard stars on each night. Reduction was performed using the IRAF\footnote{IRAF is distributed by the National Optical Astronomy Observatories, operated by AURA, Inc., under cooperative agreement with the NSF.} package and followed the standard procedures, consisting in bias image subtraction, flat-fielding, wavelength and flux calibration. Figure 2 displays sample spectra for two objects, including the main diagnostic lines.
Emission line fluxes were calculated by adopting gaussian profiles; a gaussian deblending routine was also used when necessary. Table 2, available electronically at the CDS, contains extinction-corrected fluxes in the H$\beta$ = 100 scale for the objects of our sample. See section 7.1 for a discussion on the accuracy of the results.

       \begin{table*}
        \caption{ Log of the Observations}
	\label{table1}
	\centering
	 \begin{tabular}{llll||llll}
            \hline\hline
            &Source   & Date & Observatory  &&Source   & Date & Observatory  \\
            \hline
1&K6-6& 06/25/2003 & LNA & 32&Hen  2-379& 06/24/2003 & LNA \\   
2&SS73  117 & 06/26/2003 & LNA&33&V2905  Sgr& 06/26/2003 & LNA \\ 
3&SS73  141 & 06/26/2003 &LNA&34&V4018  Sgr & 06/07/2002 & LNA \\ 
 4&Th  3-29 & 06/23/2003 &LNA & 35&SS73  122 & 06/16/2002 & LNA\\
5&H 1-25 & 06/13/2002 &LNA &36&Ap  1-8 & 06/13/2002 & LNA \\ 	  
6&Th  3-17 & 06/23/2003 &LNA & 37&V2506  Sgr & 06/25/2003 & LNA \\
7&Hen  3-1410& 06/24/2003& LNA&38&Pt  1 & 06/15/2002 & LNA \\  	     
8&AS  210  & 08/10/2002 & ESO &39&H  2-38 & 06/07/2002 & LNA \\ 
&& 06/12/2002 & LNA& 40&V2756  Sgr & 06/26/2003 & LNA \\ 	 
9&H  2-5 & 06/12/2002 & LNA& 41&SS73  129 & 06/16/2002 & LNA\\ 	   
10&H  1-36 & 06/13/2002 & LNA&42&HD  319167 & 10/09/2002 & ESO \\
11&UKS  Ce 1 & 06/12/2002 & LNA&43&V4074  Sgr & 06/07/2002 & LNA \\
12&HK  Sco& 06/14/2002 & LNA&44&V3804  Sgr & 06/14/2002 & LNA \\ 	   
13&CL  Sco & 06/07/2002 & LNA&45&Hen  3-1342 & 06/14/2002 & LNA \\ 
14&WSTB 19W032 & 06/16/2002 & LNA&46&AS 255 & 06/13/2002 & LNA \\ 	   
15&Y  CrA & 06/13/2002 & LNA&47&AS 269 & 10/08/2002 & ESO \\ 	  
16&Hen  2-171 & 06/07/2002 & LNA&48&SS 73 96 & 06/16/2002 & LNA \\
17&AE  Ara & 06/12/2002 & LNA&49& H 2-34 & 06/16/2002 & LNA \\ 	     
18&V343 Ser & 06/07/2002 & LNA&50&SS73 71& 10/10/2002 & ESO \\ 	     
19&FN  Sgr  & 06/07/2002 & LNA&51&CD 43 -14304 & 06/23-24/2003 & LNA\\ 
&& 06/23/2003 & LNA&52&Hen 3-1761 & 06/23/2003 & LNA \\ 	   
20&MWC  960 & 06/14/2002 & LNA&53& R Arq & 06/23/2003 & LNA  \\ 	     
21&RT  Ser & 06/07/2002 & LNA&54 & RR Tel & 06/23/2003 & LNA  \\ 	     
22&MaC  1-9 & 06/16/2002 & LNA&55 & V919 Sgr & 06/23/2003 & LNA \\
23&UU  Ser & 06/25/2003 & LNA&56 & Hen 3-863 & 06/24/2003 & LNA \\ 
24&V2601  Sgr & 06/14/2002 & LNA& 57 & LT Del & 06/24/2003 & LNA \\ 
25&V3811 Sgr  & 06/26/2003 & LNA& 58 & WRAY 16 377 & 06/24/2003 & LNA \\
26&ALS 2 & 06/24/2003 & LNA&59 & Bl 3-6 & 06/25/2003 & LNA \\ 	      
27&V4141  Sgr& 06/16/2002 & LNA&60 & SS73 29 & 06/25/2003 & LNA \\
28&V2416  Sgr & 06/07/2002 & LNA& 61 & AG Peg & 06/26/2003 & LNA \\ 
29&M  1-21& 06/26/2003 & LNA&62 & AS 327 & 06/26/2003 & LNA \\ 	    
30&Hen  3-1591& 06/14/2002 & LNA &63& FG Ser & 06/26/2003 & LNA \\ 
& & 10/13/2002 & ESO & 64 & PU Vul & 06/26/2003 & LNA \\ 	     
31&V2116  Oph & 06/12/2002 & LNA& &&& \\ 	  
            \hline
         \end{tabular}
 \end{table*}

\begin{table*}
      \caption{ Reddening corrected line fluxes (H$\beta$=100) }
         \label{table2}
     \centering
         \begin{tabular}{lllllllllll}
            \hline\hline
Source   & H3835 &[NeIII]$\lambda$3869  & [NeIII]$\lambda$3968&H$\delta$ &H$\gamma$ & [OIII]$\lambda$4363 &  HeII$\lambda$4686 & [ArIII]$\lambda$4711+&H$\beta$ & .. \\
& && && & &  &[NeIV]$\lambda$4714 & &  \\
        \hline\hline
     CL Sco& 8.465&       34.57&	    29.19 &  29.06 &  49.00&	  40.06 &     14.56&	 2.585&       100.0 &..\\
       FN Sgr& 5.092&       7.057&	    17.56 &  27.74 &  46.08&	  9.466 &     94.53&	 9.759&       100.0 &..\\
       H2-38& \nodata&       78.17&	    35.10 &  26.83 &  47.25&	  113.2 &     78.45&	 3.484&       100.0 &..\\
    Hen 2-171& 2.571&       49.84&	    31.09 &  22.67 &  45.00&	  44.57 &     85.27&	 6.072&       100.0 &..\\
       RT Ser& \nodata&       \nodata&	    12.67 &  20.89 &  44.25&	  7.750 &     118.8&	 1.416&       100.0 &..\\
       V2416 Sgr& \nodata&       1.423&	    0.005 &  18.19 &  38.11&	  4.308 &     80.06&	 \nodata&       100.0 &..\\
        V343 Ser& \nodata&       \nodata&	    0.023 &  20.72 &  39.03&	  2.838 &     49.37&	 5.384&       100.0 &..\\
       V4018 Sgr& 14.55&       3.301&	    19.37 &  25.45 &  44.77&	  6.526 &     49.24&	 3.362&       100.0 &..\\
       AE Ara& 9.030&       15.79&	    24.07 &  32.43 &  43.61&	  15.36 &     14.83&	 2.243&       100.0 &..\\
        H2-5& \nodata&       \nodata&	    \nodata &  20.45 &  41.17&	  24.08 &     101.7&	 \nodata&       100.0 &..\\
       AS 210& 5.381&       12.51&	    18.61 &  24.35 &  44.53&	  16.78 &     67.09&	 1.126&       100.0 &..\\
       Ap 1-8& \nodata&       \nodata&	    7.155 &  18.99 &  41.77&	  1.692 &     77.99&	 \nodata&       100.0 &..\\
..& ..&       ..&	    .. &  ..&  ..&	  ..&  ..&	 ..&       .. &..\\
        \hline
         \end{tabular}
 
   \end{table*}


\section{Reddening correction}

In the low density limit, Balmer ratios can be used to derive interstellar extinction by comparing line ratios predicted by the recombination theory with the observed values, as described by Osterbrock (1989). It was already noticed that in symbiotic stars the Balmer ratios are far from the expected values resulting from interstellar extinction only Costa \& de Freitas Pacheco (1994). These deviations from Case-B can be attributed to self-absorption effects, as described by \cite{netzer} for AGNs. Considering that symbiotic nebulae are very dense, optical depth effects should be present, therefore we adopted the \cite{netzer} and \cite{almog} results for derivation of the reddening correction and HeI abundances.

Reddening correction was derived according to the procedure described by Guti\'errez-Moreno \& Moreno (1996). This method determines simultaneously reddening and optical depth in H$\alpha$ using  the Balmer decrement values computed by Netzer (1975) for different electron densities and different optical depths at Ly$\alpha$ ($\tau_\alpha$), under conditions of self-absorption. When using this procedure, we have adopted the interestellar absorption curve from Cardelli et al. (1989), and used Netzer's (1975) Balmer decrement values for log N$_e$ = 8. 

Another direct graphical approach to the problem has been proposed (see for example \cite{pereira95}, \cite{costa})
, resulting in similar values for the reddening correction. Figure 1 shows this method, displaying H$\alpha/H\beta$ vs. H$\beta/H\gamma$ and H$\alpha/H\beta$ vs.H$\beta/H\delta$ for the objects of our sample. The curves are parametrized in optical depth in H$\alpha$ ($\tau_{H\alpha}$), for $\tau_\alpha$=10$^6$ which correspond to optically thicker nebulae.  The straight line indicates the reddening vector in both panels. 
Basically, the method consists in the determination of the H$\alpha/H\beta$ value that corresponds to the intersection of the object's reddening vector (paralel to the reddening vector in fig. 1) and the ($\tau_\alpha$) curve. However, the graphical nature of this method can potentially lead to errors due to the limited sampling of the Balmer decrement computed by \cite{netzer}, as can be seen in the figure.

In view of these limitations we adopted the analytical procedure described by Gutierrez-Moreno \& Moreno (1996). A few objects in our sample have reddening values from the 2200\AA$ $ feature listed in Table 3, derived from IUE. We did not use these values in order to retain the same criteira for the reddening determination to the whole sample. 

Nevertheless it should be noted that derivation of reddening correction for symbiotic stars can be performed using different techniques using their spectra, depending on the available spectral range and resolution, as pointed out by Mikolajewska et al. (1997), and the resulting values are usually quite different and dependent of the adopted method. The E(B-V) calculated for our sample and extracted from the literature are listed in Table 3.  

Fig. 2 displays spectra of two objects of our sample. The most important diagnostic lines are identified.

\begin{figure*}
   \centering
   \includegraphics[width=12cm]{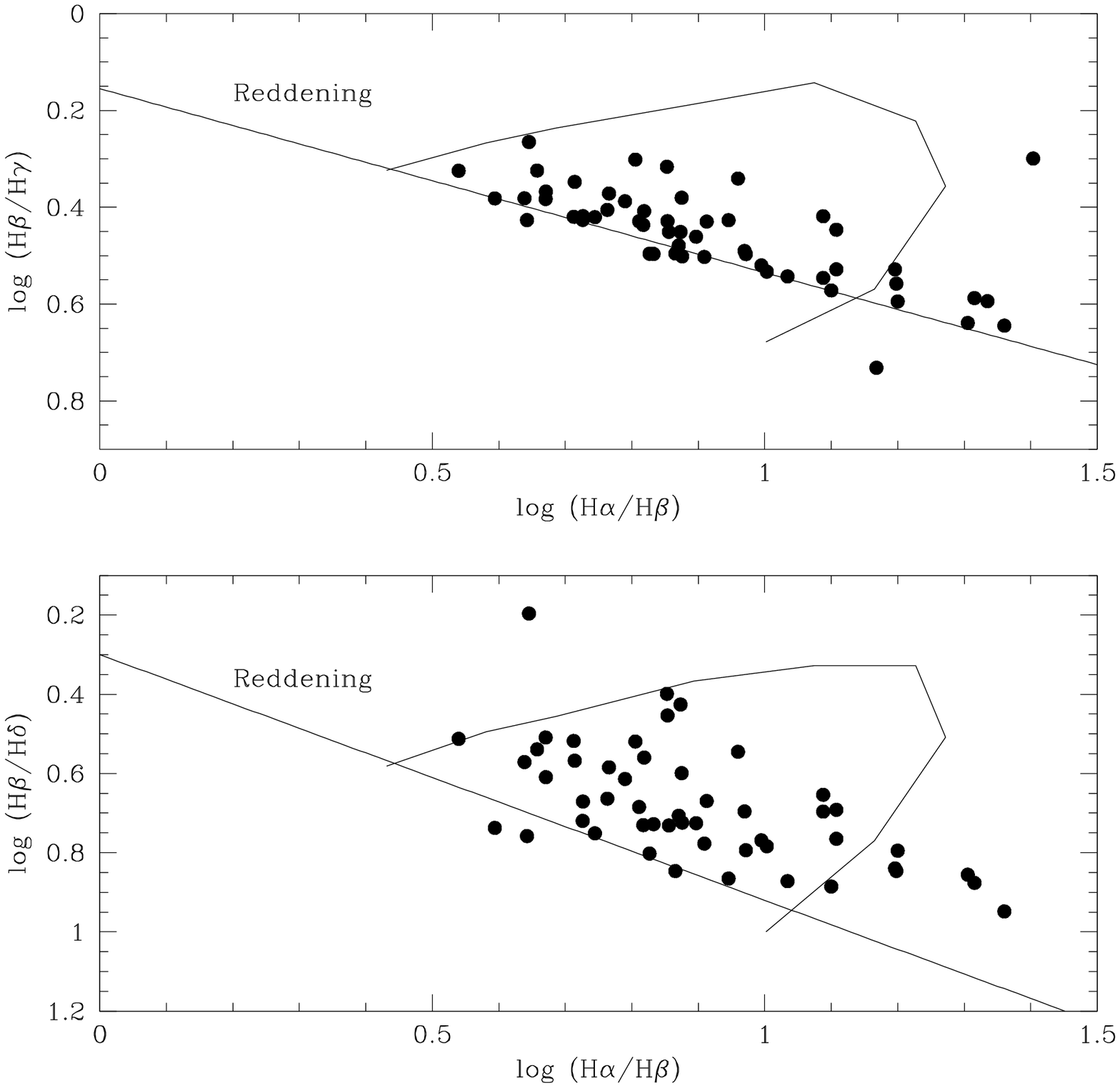}
       \caption{H$\alpha/H\beta$ vs. H$\beta/H\gamma$ and H$\alpha/H\beta$ vs.H$\beta/H\delta$ for the objects of our sample. The figure has the same format as used by Netzer (1975), the curves are parametrized in $\tau_{H\alpha}$ values. We have used only  $\tau_\alpha$=10$^6$ which correspond to optically thicker nebulae. For values of $\tau_\alpha$ and $\tau_{H\alpha}$ see Netzer's (1975) Table 1. }
   \end{figure*}

\begin{figure*}
   \centering
\includegraphics[width=12cm]{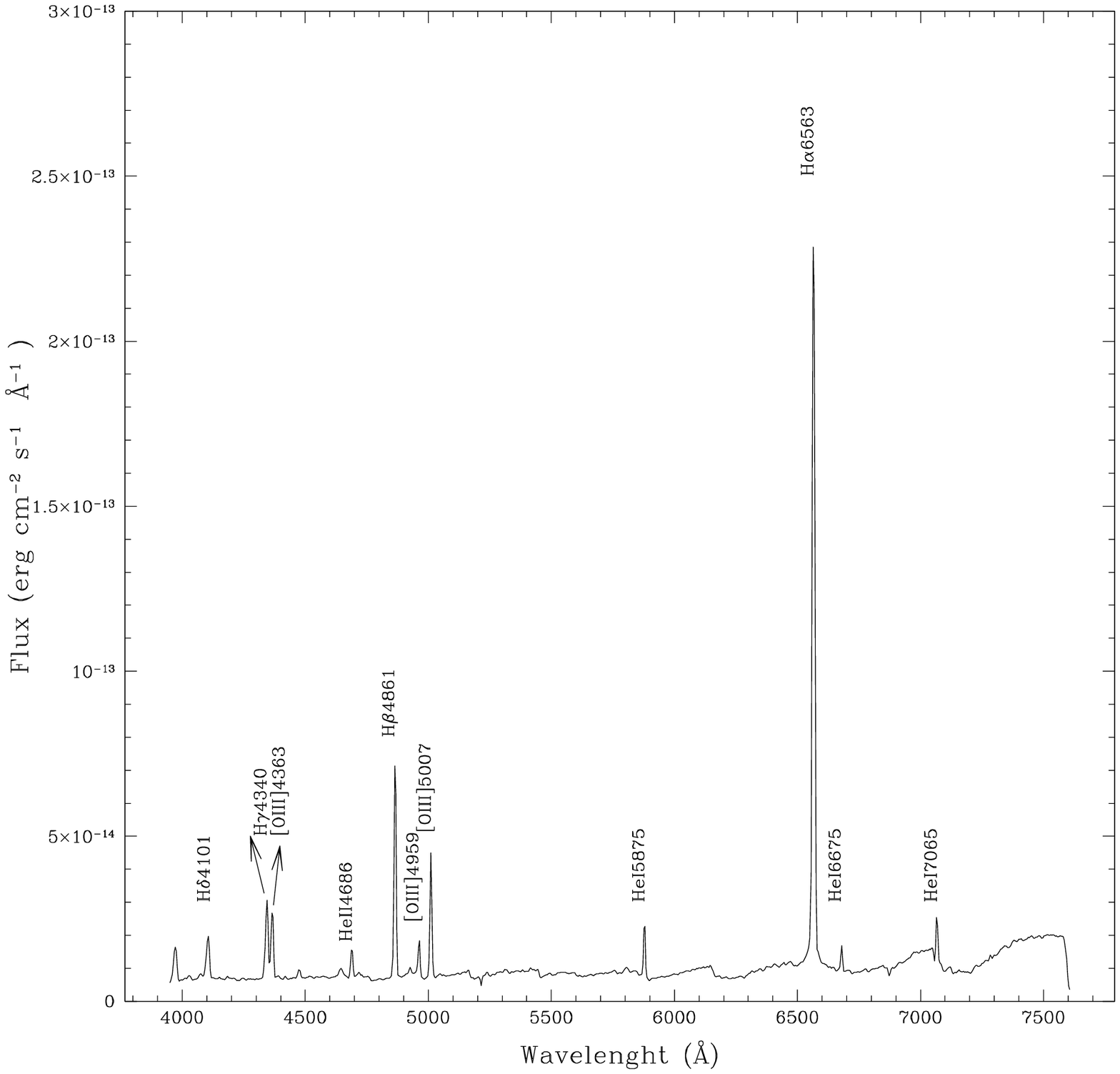}
   \includegraphics[width=12cm]{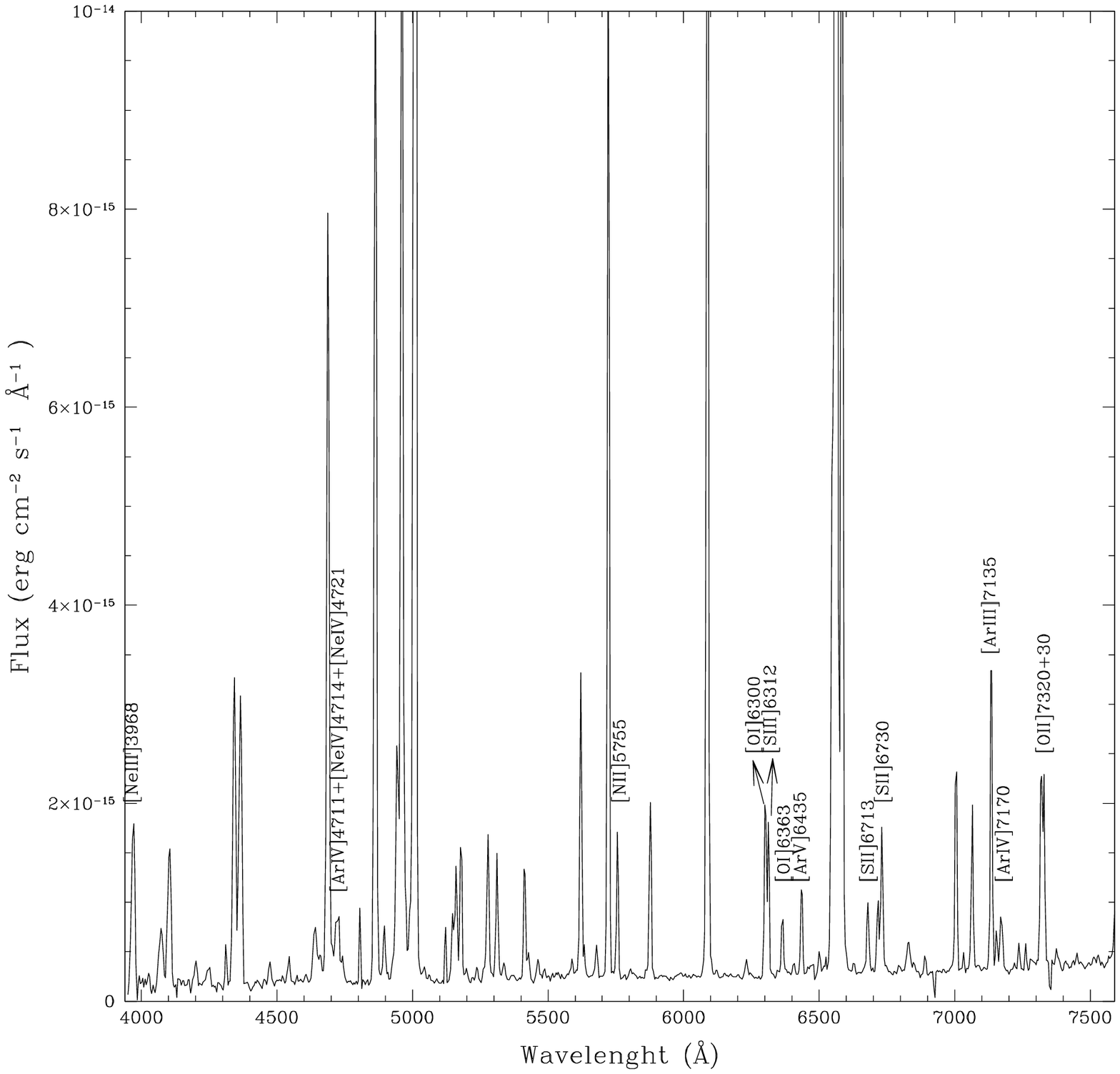}
   
    \caption{Diagnostic lines in the spectra of Hen 2-171 (lower) and CL Sco (upper).}
   \end{figure*}

\section {Physical conditions}

The usual forbidden line ratios used to derive physical parameters of emission nebulae do not allow a unique electron temperature determination since the symbiotic nebula should have an important density stratification. However the derivation of the ionic concentrations in the ionized gas requires a previous knowledge of the electron temperature and density. From optical spectra these parameters can usually be estimated from line intensity ratios like, among others:
\begin{eqnarray}
R([OIII]) = \frac{\lambda5007 + \lambda4959}{\lambda4363}\\
R([NII]) = \frac{\lambda6584 + \lambda6548}{\lambda5755}
\end{eqnarray}
\begin{equation}
R([SII]) = \frac{\lambda6717}{\lambda6730}
\end{equation}

In view of the high density of symbiotic nebulae is difficult to use the R(SII) to estimate the electron density, as this relation produce unique results for N$_e \leq 10^{5}$ (\cite{osterbrok}). We estimated a lower limit for the density from the R([OIII]) and R([NII]) ratios assuming T$_e$ $\approx$ 12000 K, which is an acceptable value for symbiotics (\cite{nussbaumer}), and adopted two different density zones, namely N$_e$[OIII] and N$_e$[NII]; for those objects where both ratios were available. Otherwise a single zone was assumed. Fig. 3 displays an example of the behavior of R([OIII]) and R([NII]) for Hen 2-171. 

R([NII]) and R([OIII]) characterize two distinct regions where the relations above are sensitive to density, resulting in lower limits for this parameter. This picture is clearly an oversimplification of the true situation, however, high resolution IUE spectra of V 1016 Cyg, suggests an interpretation not inconsistent with such an approximation (\cite{deuel}). Narrow and broad components suggesting at least two regions with distinct densities are also seen in Coud\'e spectra of HM Sge (\cite{stauffer}). We consider that in the R([NII]) region the main ionic species are N$^+$, O$^+$, S$^+$, while in the R([OIII]) region species of higher excitation like O$^{+2}$, S$^{+2}$, Ne$^{+2}$, Ar$^{+2}$, Ar$^{+3}$ are dominant. Only in AS 210, RR Tel, PN H 1-36, PN H 2-38 and Hen 2-171  both density regions were used. For these objects R([NII]) lead to density values (in cm$^{-3}$) of respectively : 2.6*10$^5$, 3.38*10$^5$, 1.03*10$^5$, 3.38*10$^5$, 1.*10$^5$ . The derived values for N$_e$ are listed in Table 3. 


\begin{figure*}
   \centering

   \includegraphics[width=12cm]{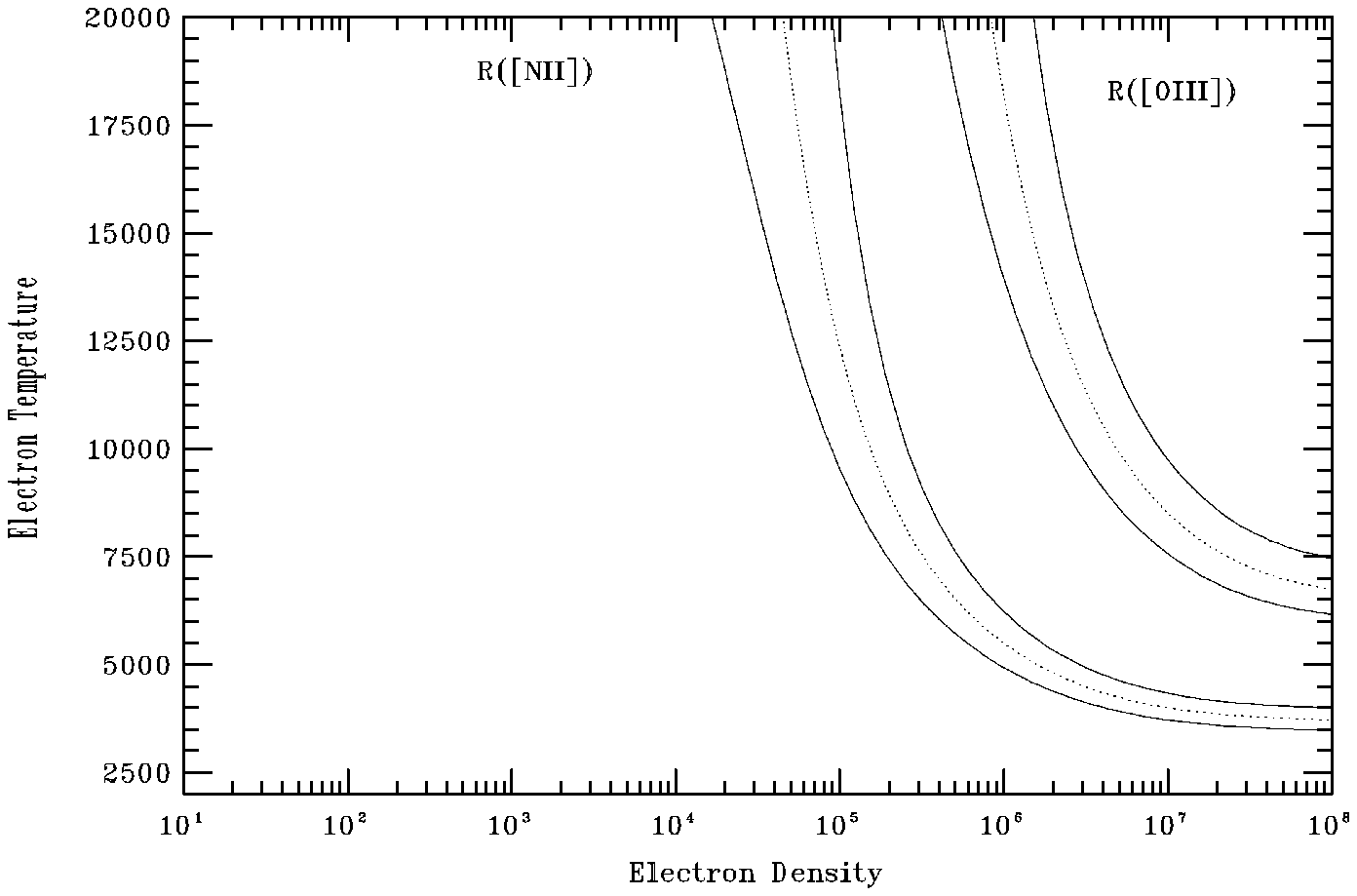}
   
    \caption{Behavior of R([OIII]) and R([NII]) relations for Hen 2-171. The solid lines represent an uncertainty of 2$\sigma$ with respect to the calculated value}
   \end{figure*}


\begin{table*}
      \begin{minipage}[t]{\columnwidth}
  \caption{N$_e$ (cm$^{-3}$), E(B-V) and $\tau_{H\alpha}$ derived for our sample}
         \label{table3}
     \centering
     \renewcommand{\footnoterule}{}
       \begin{tabular}{llllll||llllll}
            \hline\hline
&Source & E(B-V) & $\tau_{H\alpha}$ & N$_e$[OIII]&Other E(B-V)&&Source & E(B-V) & $\tau_{H\alpha}$ & N$_e$[OIII]&Other E(B-V) \\
\hline
1&K6-6&2.23 & 15.4 & 5.29*10$^{6}$& &35&SS73  122 &0.84& 9.9& 4.38*10$^{6}$&0.92 \footnote{Pereira (1995)}, 1.3$^{f}$ \\  
3&SS73  141$^{k}$ & 0.55 &1.3& 1.0*10$^{8}$& &36&Ap  1-8 & 1.04& 3.6& 4.12*10$^{6}$ &0.6$^{f}$ \\  
4&Th  3-29 & 1.77$^{b}$ &20.5 & 1.19*10$^{7}$&2.7$^{f}$&37&V2506  Sgr$^{k}$ & 0.64 &3.9& 1.0*10$^{8}$  &0.5$^{f}$\\  
5&H 1-25 & 2.54\footnote{using H$\alpha$ \& H$\gamma$}&3.6&1.44*10$^{7}$&&38&Pt  1 & 0.95&8.9&4.17*10$^{7}$  &\\  
7&Hen  3-1410& 1.44$^{b}$&9.7& 8.73*10$^{6}$&&39&H  2-38 & 0.66 &15&6.4*10$^{6}$ &0.51$^{d}$, 1.2$^{f}$\\  
8&AS  210 &0.49&4.3& 3.87*10$^{6}$ & $\leq$ 0.5\footnote{Schmid \& Nussbaumer (1993)}, 0.30 $^{d}$& 40 & V2756  Sgr & 0.32 & 8.75 & 1.05*10$^{7}$ & 0.0$^{f}$ \\ 
9&H  2-5 & 1.28 &12.5&2.69*10$^{7}$&1.8$^{f}$&41&SS73  129$^{k}$ & 0.78& 3.4 & 1.0*10$^{8}$ &1.6$^{f}$\\  
10&H  1-36 & 0.51&1.4&4.92*10$^{5}$ &0.71\footnote{Pereira et al. (1998)}&42&HD  319167 & 0.66&3.3&3.67*10$^{7}$  &1.0$^{f}$\\  
12&HK  Sco& 0.59&6.1 & 4.16*10$^{7}$&&44&V3804  Sgr & 0.25& 7.9& 5.64*10$^{6}$  &\\  
13&CL  Sco & 0.22&6.4&3.67*10$^{7}$ &0.2\footnote{Michalitsianos et al. (1982)}&45&Hen  3-1342 &0.63& 4.1&2.66*10$^{6}$  &\\ 
15&Y  CrA & 0.33&17.1&1.96*10$^{7}$&0.23$^{l}$&46&AS 255$^{k}$ & 0.68&8.0&1.0*10$^{8}$  &\\ 
16&Hen  2-171 & 0.59& 2.9& 2.66*10$^{6}$&0.58&47&AS 269 &2.08$^{b}$&2.7& 1.0*10$^{8}$ &2.4\footnote{Mikolajewska et al. (1997)}\\ 
17&AE  Ara & 0.14&5.6& 1.53*10$^{7}$  &0.5$^{f}$&48&SS 73 96&1.48&10.1&1.0*10$^{8}$  &\\ 
18&V343 Ser & 1.18&8.5& 1.19*10$^{7}$ & &49& H 2-34 & 1.29&11.7&1.05*10$^{7}$  &\\ 
19&FN  Sgr\footnote{N$_{e}$ assumed minimum value}  & 0.22& 2.9& 1.0*10$^{8}$ & 0.6$^{j}$&50&SS73 71& 0.34&14.6&8.2*10$^{6}$  & 0.42\footnote{Pereira et al. 2002} \\ 
20&MWC  960 & 0.84&4.9& 1.63*10$^{7}$ &0.7$^{j}$ &51&CD 43 -14304$^{j}$ & 0.76&3.0&1.0*10$^{8}$ &$\leq$ 0.2$^{c}$\\ 
21&RT  Ser$^{g}$ & 1.08& 10.8& 1.0*10$^{8}$&&52&Hen 3-1761 &0.15&12.6&1.4*10$^{7}$  &\\ 
22&MaC  1-9$^{k}$ & 1.14&4.72&1.0*10$^{8}$  &1.2$^{f}$&53& R Arq &0\footnote{Simon (2003)} &0&2.6*10$^{5}$  &0.08$^{l}$\\ 
23&UU  Ser & 0.63&4.55&1.96*10$^{7}$ &1.2$^{f}$&54 & RR Tel & 0.05&4.3&4.67*10$^{6}$  &0.09$^{l}$\\ 
24&V2601  Sgr &0.39&4.3&2.86*10$^{7}$&&55 & V919 Sgr$^{g}$ &0.38&10.3&1*10$^{8}$  &\\ 
26&ALS  2$^{g}$ & 0.94&2.1& 1.0*10$^{8}$&1.0$^{f}$&56 & Hen 3-863 & 0.19&7.4&1.0*10$^{8}$  &\\ 
27&V4141  Sgr& 1.12&9.6 & 6.79*10$^{6}$ &1.2$^{f}$&57 & LT Del$^{g}$ & 0.41&\nodata&1.0*10$^{8}$  &\\ 
28&V2416  Sgr & 1.69&14.0& 6.06*10$^{7}$  &2.5$^{f}$ &58 & WRAY 16 377 & 0.83&1.15&2.09*10$^{7}$  &\\ 
29&M  1-21& 0.88&5.5& 1.0*10$^{8}$ &1.0$^{f}$&60 & SS73 29 & 0.50&\nodata&1.53*10$^{7}$  &1.0\footnote{Munari \& Buson (1994)}\\ 
30&Hen  3-1591& 0.06&19.0& 1.1*10$^{7}$ &&61 & AG Peg\footnote{without diagnostic lines, N$_e$=10$^8$}  & 0.36&6.5&1.0*10$^{8}$  &0.12\footnote{Nussbaumer et al. (1988)}\\ 
32&Hen  2-379& 0.17&16.2&1.0*10$^{8}$ &&62 & AS 327$^{k}$ & 0.86&5.4&1.0*10$^{8}$  &1.1$^{f}$\\ 
33&V2905  Sgr& 0.43& 3.8&3.24*10$^{7}$  &&63& FG Ser & 0.78&14.7&4.72*10$^{7}$ &0.82\footnote{Guti\'errez-Moreno et al. (1992)}\\ 
34&V4018  Sgr & 0.44&3.9&4.16*10$^{7}$ &0.4\footnote{Munari \& Buson (1993)}&64 & PU Vul &0.29&6.8&1.53*10$^{7}$  &\\ 
\hline
\end{tabular}
\end{minipage}
\end{table*}

\section{Helium abundance}

Helium  abundance is a key parameter to characterize chemical evolution either of stars and galaxies. As discussed by Clegg (1987), line formation in HeI is complex in view of the metastability of the lowest triplet level, which can cause some lines to become optically thick. In particular, the collisional enhancement of HeI lines from the metastable 2$^3$S level is an important issue that cannot be ruled out, as indicated by Kingdon \& Ferland (1995). This is a specially controversial subject in symbiotic stars in view of their high nebular densities.

Usually helium abundance is expressed relative to hydrogen. As discussed in section 4, in high density nebulae self-absorption effects are present in the Balmer series. In order to minimize these effects on the helium abundance, Schmid \& Schild (1990) used the ratio between HeII and higher excitation Balmer lines when deriving the final helium abundance. Here we adopted a similar procedure, taking H$\gamma$ as our reference line. It was chosen as a compromise solution between lower lines, more affected by self absorption effects, and higher, weaker, lines. The optical depth in this line can be scaled to the value of $\tau_{H\alpha}$. Using the optical depth in H$\alpha$, $\tau_{H\alpha}$, which is derived from our reddening calculation, the optical depth in H$\gamma$, $\tau_{H\gamma}$ is 0.07*$\tau_{H\alpha}$. In the cases that  $\tau_{H\gamma}$ $\leq$ 0.5 we would expect that self-absorption effects are negligible and no large errors are being committed.

The HeI abundances were computed considering the possibility of large collisional excitation and self absorption effects, following the procedure described in details by Costa \& de Freitas Pacheco (1994). In the present work we derived the abundance from lines $\lambda$5876 and $\lambda$7065, weighted by their intensities.  Line $\lambda$6678 was not used to compute HeI abundance because we have detected that in many objects it is placed over the wings of H$\alpha$, resulting in overestimated fluxes. Line  $\lambda$7065 was also used to estimate the optical depth in $\lambda$3889 ($\tau_{(\lambda3889)}$) since this optical depth is required to derive emissivities calculated by \cite{almog} for HeI lines. 
The concentration of He$^+$ can consequently be obtained from the relation (de Freitas Pacheco \& Costa, 1992):

\begin{eqnarray}
\frac{n(He+)}{n(H)} = \frac{5.9}{E_{\lambda}(\tau)}\frac{I_\lambda}{I_\gamma}
\end{eqnarray}

where I$_{\lambda}/I_{\gamma}$ is the line intensity ratio between the considered HeI line and H$\gamma$ and E$_{\lambda}$($\tau$) is the emissivity calculated by \cite{almog} ( see their table 2) as a function of $\tau(\lambda3889$). As pointed by de Freitas Pacheco \& Costa (1992), the intensity of the HeI$\lambda$7065 line in RR Tel is too high and produce unrealistic values for $\tau(\lambda3889$) and therefore higher HeI abundance. Table 4 shows the resulting He$^+$ abundances together with the He$^{+2}$ derived using the $\lambda$4686 line.

   \begin{table*}
      \caption{He abundances}
         \label{table3}
	 \centering
         \begin{tabular}{lllll||lllll}
            \hline\hline
             &Source   & He I & He II & He &&Source   & He I & He II & He \\
              \hline
1& K6-6& 0.232 & 0.033 & 0.265&  35&SS73  122 & 0.177 & 0.048&0.225\\  
5&H 1-25 & \nodata &\nodata & \nodata & 36&Ap  1-8 & 0.113 & 0.068 & 0.181 \\  
7&Hen  3-1410& 0.191&0.047 &0.238&  37&V2506  Sgr & \nodata &\nodata& \nodata \\8&AS  210 & 0.029 &0.058 & 0.087 &  38&Pt  1 & \nodata & \nodata & \nodata \\  
10&H  1-36 &0.057 &0.054 &0.111&  39&H  2-38 & 0.129 & 0.068 & 0.197 \\  
12&HK  Sco& 0.083 & 0.088 & 0.171 &40&V2756  Sgr & 0.177 & 0.071 & 0.248 \\
13&CL  Sco & 0.110 &0.013 & 0.123 &  42&HD  319167 & 0.093 &0.009 & 0.102 \\  
15&Y  CrA & 0.115 & 0.017 & 0.132 &  44&V3804  Sgr & 0.207 & 0.014 & 0.221 \\  
16&Hen  2-171 & 0.063 & 0.074 & 0.137& 45&Hen  3-1342 & 0.167 & 0.053 & 0.220\\ 17&AE  Ara & 0.141 & 0.013 & 0.54 & 46& AS 255 & 0.179 & 0.092 & 0.271 \\ 
18&V343 Ser & \nodata & \nodata&\nodata&50&SS 73 71 & 0.182 & 0.038 & 0.220 \\ 
19&FN  Sgr  & 0.099 & 0.082 & 0.181 & 51&CD 43 -14304 & 0.134 & 0.059 & 0.193 \\ 20&MWC  960 & 0.127 & 0.061 & 0.188&52& Hen 3-1761 & 0.217 & 0.048 & 0.265 \\ 
21&RT  Ser & \nodata& \nodata & \nodata& 53& RR Tel &\nodata &\nodata&\nodata\\ 23&UU  Ser & 0.102 & 0.048 & 0.15 & 55 & V919 Sgr & 0.126 & 0.041 & 0.167 \\ 
24&V2601  Sgr & 0.122 & 0.025 & 0.147&  56 & Hen 3-863 & 0.098 & 0.052 & 0.150 \\
26&ALS  2 & 0.096& 0.050 & 0.146& 57 & LT Del & 0.113 & 0.020  & 0.133 \\ 
27&V4141  Sgr& 0.179 & 0.006 & 0.185&58 & WRAY 16 377 & \nodata&\nodata&\nodata \\ 
28&V2416  Sgr & 0.084 & 0.070 & 0.154& 61 & AG Peg & 0.128 & 0.087 & 0.215 \\ 
29&M  1-21& 0.093 & 0.040 & 0.133&62& AS 327 & 0.163 & 0.083 & 0.246 \\ 
33&V2905 Sgr& 0.168 & 0.006 & 0.174&64 & PU Vul & 0.083 & 0.063 & 0.146 \\ 
34&V4018  Sgr & 0.121 & 0.043 & 0.164 & &&&& \\ 
					
	     \hline
         \end{tabular}
    \end{table*}

\section{Abundances of heavy elements}

In order to minimize the effects of self-absorption in Balmer lines on the derived abundances, we calculated elemental ratios with respect to oxygen, since they can be derived directly from forbidden transitions. In the R(NII) region we derive the nitrogen to oxygen ratio from the relation

\begin{equation}
\frac{N}{O} \approx \frac{N^+}{O^+}
\end{equation}

In the R([OIII]) region, the relative abundances of neon and argon with respect to oxygen can be obtained from Schmid \& Schild (1990):

\begin{equation}
\frac{Ne}{O} \approx \frac{Ne^{+2}}{O^{+2}}
\end{equation}

\begin{equation}
\frac{Ar}{O} \approx \frac{Ar^{+2}+Ar^{+3}}{O^{+2}}
\end{equation}

To derive the ionic concentrations we have adopted a five level atom model, including collisional excitation and desexcitation and radiative transitions in the statistical balance equations. The model is described by Shaw \& Dufour (1994), and the relevant atomic data used can be retrieved directly from IRAF $nebular$ package 

Relative elemental abundances of heavier elements can be derived from emission lines which originate from a common region. In dense nebulae, the ionization structure of helium is particularly well defined by the Str\"{o}mgren radii of He$^{+2}$ and H$^+$. It is therefore appropriate to compare ions which coincide with either the  He$^{+2}$ or  He$^{+}$ regions. With our spectral resolution, we can identify in some objects two different density regions, one for the N$^+$ and the other for O$^{+2}$, therefore as was pointed earlier we calculate relative elemental abundances for ions formed in the same density region. 

The  $N^+$ abundance was derived from  $\lambda$6548+84 and/or $\lambda$5755, and $O^+$ was derived from  $\lambda$7320+30 lines. For the O$^{+2}$ abundance we have used a mean value between [OIII]$\lambda$4363, [OIII]$\lambda$4959 and [OIII]$\lambda$5007, weighted by their intensities, and for Ne$^{+2}$ we have used the [NeIII]$\lambda$3869 line when available, or [NeIII]$\lambda$3968. In almost all the sample, the dependence of the Ne/O relation to N$_c$ (critical density) for the used lines is not very large, because the densities are similar or greater than N$_c$ (\cite{schmidt}). The [ArIII]$\lambda$7136 and [ArIV]$\lambda$4740 lines were used together with [OIII] abundance for the Ar/O determination; the [ArIII]$\lambda$4711 line cannot be used because with our spectral resolution was blended with [NeIV]$\lambda$4714. The ionic abundances where calculated with the task \textit{nebular/abund} of IRAF and are listed in Table 5. They were calculated adopting T$_e$=12000 K and taking N$_e$ from table 3. The derived relative elemental abundances are listed in Table 6.

\subsection{Precision of the abundance determination}

\begin{figure}
   \centering
   \includegraphics[width=8.8cm]{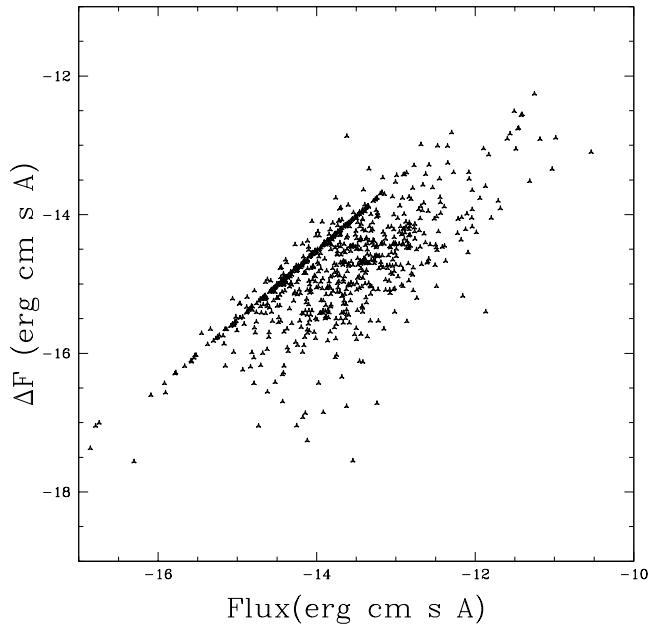}
    \caption{Flux mean values vs. standard deviation ($\Delta$F) from all the lines for each object of the sample}
    \label{fig3}
   \end{figure}

Abundance determination involves many sources of errors like accuracy of the fluxes, limited precision in data reduction, uncertainties in reddening and the derived physical parameters, as well as in the adopted methodology to derive abundances. We tested the sensitivity of the abundances to the line fluxes for each object by calculating mean and standard deviation for all the flux measurements of each spectral line for each object. The results can be seen in fig. \ref{fig3}, where $\Delta$F represents the standard deviation for each line. The points that form a straight line correspond to those lines for which we have only one measurement; for them we adopt an error of 30\% for the lines weaker than (1/3)H$\beta$ and 10\% for the stronger lines (respect to H$\beta$=100). The flux dispersions result in abundance dispersions that have a mean value of 0.2 dex for stronger lines like [OIII], and 0.4 dex for weaker lines like [ArIV],[NeIII], etc. These uncertainties are typical in this type of calculations (e.g. \cite{escudero}).

 We also tested the dependence of the derived abundances on E(B-V), the most uncertain parameter along the whole process. As described in section 4, we calculate E(B-V) using the method described by \cite{moreno1}. Just to check the dependence of the derived relative abundances to the reddening, we consirered an uncertainty of 50\% in E(B-V) and rederived the electron densities and abundances, which resulted in upper and lower limits of precision for our results. These limits are displayed in figure 5 as a mean error bar for the results. It can be seen that even such a large error in reddening does not affect our main conclusions.


%

\section{Results and discussion}

Our sample contains 54 objects for which ionic abundances have been derived. For many of them, these are the first abundances published. These abundances are similar to those obtained for nebulae that present similar spectroscopic behavior as planetary nebulae. 

To perform the chemical diagnostics we have chosen emission lines from elements with similar ionization potentials, in order to minimize the effects of density gradients. We have obtained abundances ratios from collisionally excited lines, namely N/O, Ne/O and Ar/O. 

 We present the ionic concentrations for O, N, Ar, S, He and Ne. The relative elemental abundances show the composition of the interstellar medium in the time of progenitor formation, and the N and He abundances reflect the evolutionary state of the stars, showing that this class of objects have experienced dredge up episodes, may be up to the second one. Also the He and N abundances are comparable with progenitors between 0.6 to 1.5 M$\odot$.

Fig. 5 shows the relation between N/O and He/H; as nitrogen and helium are nucleosynthesis products and related to the mass spectrum of the progenitors, they are expected to be correlated; and this diagram is important to the diagnostic of abundances. The figure combines data from symbiotics with other from planetary nebulae ( Escudero \& Costa 2001, Escudero et al. 2004, Exter et al. 2004). It can be seen that the objects in this figure are distributed in two groups: a larger, upper group for which the planetary nebulae sample define a reasonably well defined correlation between log(He/H) and log(N/O), in the sense that helium-rich objects are usually nitrogen-rich. The symbiotics fit into this distribution, in spite of their dispersion. This correlation reflects the mass spectrum of the progenitors. It can also be seen a small group of PNe with low log(N/O) and log(He/H) varying from -1.1 to -0.7. The same pattern appears in the results of \cite{cuisinier} for galactic PNe and can be related to objects at high Z above the galactic plane. However, the uncertainties in distances, both for symbiotics and PNe, make this point still an open question. 

The same figure also includes a model from Marigo (2001), with mixing-length parameter=1.68 and initial metallicity Z=0.019 and Y=0.273, computed to 0.1026 $\geq$ He $\geq$ 0.1387 and -0.9 $\geq$ log(N/O) $\geq$ -0.001. Clearly, the dispersion of the data is high, but the model trend agrees with the mean values and tendencies for most of the PNe sample. For some symbiotic stars, however, this agreement cannot be seen, may be due to the uncertanties in the method used to derive He abundances and/or to the Marigo's model, that was produced for isolated stars, and some discrepacies are expected with respect to binary objects like symbiotics.

Nebular abundances of symbiotics should reflect the abundances of the intermediate mass stars from which they originate, irrespective of their position in the galaxy. 
To verify this behavior we compare the mean values of the Ne/O ratio for our sample of symbiotics to other samples of disk and bulge PNe. Being $\alpha$-elements, the Ne/O ratio do not reflect the chemical evolution of the interstellar medium and should remain the same along the galactic bulge and disk. The mean values and dispersions are listed in table 7.

\begin{sidewaystable*}
\begin{minipage}[t][180mm]{\textwidth}
\caption{Ionic abundances for symbiotic stars in our sample}\label{table6}
\centering
\begin{tabular}{lccccccccccc}
	   \hline\hline
            Source&NII&OI&OII&OIII&NeIII&NeIV&SIII&ArV&ArIII&ArIV&NI\\
            \hline

       AE Ara& 3.933*10$^{-6}$& 2.157*10$^{-5}$& 3.330*10$^{-5}$& 1.470*10$^{-4}$& 2.276*10$^{-5}$& 2.665*10$^{-5}$& 1.611*10$^{-6}$& 3.417*10$^{-6}$&    \nodata&    \nodata&	 \nodata \\
       Ap 1-8& 3.417*10$^{-6}$&    \nodata&	 \nodata& 1.302*10$^{-5}$&    \nodata&    \nodata&	 \nodata& 1.787*10$^{-7}$& 1.357*10$^{-8}$& 2.018*10$^{-5}$&	 \nodata  \\
       AG Peg& 1.024*10$^{-4}$&    \nodata& 2.353*10$^{-4}$& 4.000*10$^{-4}$& 1.897*10$^{-5}$&    \nodata&	 \nodata&    \nodata&    \nodata&    \nodata&  0.05983  \\
       ALS 2& 2.478*10$^{-4}$&    \nodata& 2.622*10$^{-4}$& 7.506*10$^{-5}$& 7.235*10$^{-5}$&    \nodata&	 \nodata&    \nodata&    \nodata& 0.001206&	 \nodata  \\
       AS 210& 1.605*10$^{-6}$& 2.669*10$^{-6}$& 1.754*10$^{-6}$& 1.295*10$^{-4}$& 9.575*10$^{-6}$& 5.862*10$^{-5}$& 1.274*10$^{-6}$& 2.078*10$^{-7}$& 1.861*10$^{-7}$& 2.346*10$^{-7}$&	 \nodata  \\
       AS 255&    \nodata& 2.971*10$^{-4}$& 1.263*10$^{-4}$& 5.101*10$^{-5}$&    \nodata&    \nodata&	 \nodata&    \nodata& 6.565*10$^{-6}$& 9.061*10$^{-5}$&	 \nodata	 \\
       AS 327& 4.596*10$^{-5}$&    \nodata& 1.081*10$^{-4}$& 1.626*10$^{-4}$&    \nodata&    \nodata&	 \nodata&    \nodata&    \nodata& 8.546*10$^{-4}$&	 \nodata	 \\
      CD 43 -14304& 4.313*10$^{-6}$&    \nodata& 4.059*10$^{-5}$& 2.382*10$^{-4}$& 2.104*10$^{-5}$&    \nodata&	 \nodata&    \nodata& 2.930*10$^{-7}$& 3.232*10$^{-4}$&  0.02439	 \\
       CL Sco& 1.384*10$^{-5}$& 1.336*10$^{-5}$& \nodata& 5.814*10$^{-4}$& 9.356*10$^{-5}$& 8.066*10$^{-5}$& 1.594*10$^{-6}$& 1.273*10$^{-6}$& 1.829*10$^{-7}$& 1.450*10$^{-5}$& 0.007804	 \\
       FG Ser& 1.087*10$^{-4}$&    \nodata& 3.754*10$^{-4}$& 2.514*10$^{-4}$& 1.955*10$^{-5}$&    \nodata&	 \nodata&    \nodata&    \nodata& 5.659*10$^{-4}$&	 \nodata \\
       FN Sgr& 6.043*10$^{-5}$&    \nodata& 6.566*10$^{-5}$& 2.622*10$^{-4}$& 4.551*10$^{-5}$&    \nodata& 5.739*10$^{-6}$& 7.061*10$^{-6}$&    \nodata& 8.320*10$^{-4}$&  0.01075  \\
       H 1-25& 2.256*10$^{-4}$& 4.145*10$^{-5}$& 1.699*10$^{-6}$& 2.061*10$^{-4}$&    \nodata&    \nodata&	 \nodata&    \nodata& 8.756*10$^{-8}$& 2.783*10$^{-7}$&	 \nodata  \\
       H 1-36& 2.495*10$^{-5}$& 3.449*10$^{-5}$& 3.447*10$^{-5}$& 5.114*10$^{-4}$& 6.280*10$^{-5}$& 2.776*10$^{-4}$& 7.722*10$^{-6}$& 8.726*10$^{-7}$& 1.535*10$^{-6}$& 4.216*10$^{-6}$&   \nodata  \\
       H 2-34& 3.240*10$^{-6}$& 2.503*10$^{-5}$& 5.670*10$^{-6}$& 1.754*10$^{-4}$&    \nodata&    \nodata& 4.097*10$^{-6}$& 8.995*10$^{-7}$& 1.259*10$^{-7}$& 5.957*10$^{-6}$& 0.003656  \\
       H 2-38& 7.662*10$^{-6}$& 1.794*10$^{-5}$& 1.185*10$^{-5}$& 9.006*10$^{-4}$& 7.150*10$^{-5}$& 1.842*10$^{-4}$& 3.686*10$^{-6}$& 1.257*10$^{-6}$& 4.284*10$^{-7}$& 2.727*10$^{-6}$&	 \nodata   \\
        H 2-5&    \nodata& 9.349*10$^{-5}$&	 \nodata& 2.950*10$^{-4}$&    \nodata&    \nodata& 3.662*10$^{-6}$& 2.713*10$^{-6}$&    \nodata&    \nodata&	 \nodata   \\
    HD 319167& 1.899*10$^{-6}$& 1.640*10$^{-5}$& 4.561*10$^{-6}$& 1.838*10$^{-4}$& 2.872*10$^{-5}$& 5.349*10$^{-5}$& 1.994*10$^{-7}$& 4.307*10$^{-7}$& 5.559*10$^{-8}$& 2.977*10$^{-6}$&	 \nodata   \\
    Hen 2-171& 1.985*10$^{-5}$& 9.651*10$^{-6}$& 8.123*10$^{-6}$& 3.409*10$^{-4}$& 3.459*10$^{-5}$& 1.796*10$^{-4}$& 5.365*10$^{-6}$& 1.155*10$^{-6}$& 1.062*10$^{-6}$& 2.068*10$^{-6}$&	 \nodata	 \\
    Hen 2-379&  0.07021&    \nodata&	 \nodata& 0.001018&    \nodata& 0.001649&	 \nodata&    \nodata&    \nodata&    \nodata&	 \nodata	 \\
   Hen 3-1342&    \nodata& 1.485*10$^{-5}$& 6.465*10$^{-6}$& 1.094*10$^{-5}$&    \nodata&    \nodata& 3.548*10$^{-6}$& 4.772*10$^{-7}$& 2.308*10$^{-7}$& 5.684*10$^{-7}$&	 \nodata	 \\
   Hen 3-1410& 4.200*10$^{-6}$&    \nodata& 8.082*10$^{-6}$& 1.097*10$^{-4}$&    \nodata&    \nodata&	 \nodata& 4.579*10$^{-7}$&    \nodata&    \nodata&	 \nodata	 \\
   Hen 3-1591& 3.998*10$^{-4}$& 1.527*10$^{-4}$& 1.201*10$^{-4}$& 0.001048& 9.513*10$^{-5}$&    \nodata& 3.787*10$^{-6}$& 8.230*10$^{-7}$& 3.029*10$^{-6}$& 3.165*10$^{-5}$&	 \nodata	 \\
   Hen 3-1761& 1.113*10$^{-4}$&    \nodata& 2.676*10$^{-4}$& 3.304*10$^{-4}$& 4.483*10$^{-5}$& 1.684*10$^{-4}$&	 \nodata&    \nodata&    \nodata& 1.967*10$^{-4}$&	 \nodata	  \\
    Hen 3-863&    \nodata&    \nodata& 2.380*10$^{-4}$& 3.820*10$^{-4}$& 9.932*10$^{-6}$&    \nodata&	 \nodata&    \nodata&    \nodata& 5.861*10$^{-5}$&	0.1261	  \\
       HK Sco& 3.364*10$^{-5}$&    \nodata& 3.423*10$^{-5}$& 7.033*10$^{-5}$& 4.788*10$^{-6}$&    \nodata&	 \nodata& 3.502*10$^{-6}$&    \nodata& 6.584*10$^{-5}$&	 \nodata	  \\
        K6-6& 3.446*10$^{-6}$& 2.182*10$^{-5}$& 4.302*10$^{-6}$& 1.110*10$^{-4}$&    \nodata& 7.921*10$^{-5}$& 1.061*10$^{-6}$& 3.094*10$^{-7}$& 3.232*10$^{-8}$& 4.675*10$^{-5}$&	 \nodata	  \\
       LT Del& 4.330*10$^{-5}$&    \nodata&	 \nodata& 2.524*10$^{-5}$&    \nodata&    \nodata&	 \nodata&    \nodata&    \nodata& 6.490*10$^{-4}$&	 \nodata	\\
        M1 -21&    \nodata& 6.010*10$^{-5}$& 1.750*10$^{-5}$&    \nodata&    \nodata& 1.694*10$^{-4}$& 2.847*10$^{-6}$& 1.900*10$^{-6}$&    \nodata& 2.342*10$^{-4}$&	 \nodata	 \\
         MaC 1-9& 3.084*10$^{-5}$&    \nodata& 3.464*10$^{-5}$& 1.852*10$^{-4}$&    \nodata&    \nodata&	 \nodata&    \nodata&    \nodata&    \nodata&	 \nodata	 \\
      MWC 960& 4.429*10$^{-6}$& 3.222*10$^{-5}$& 8.880*10$^{-6}$& 2.070*10$^{-5}$&    \nodata&    \nodata& 2.093*10$^{-6}$& 6.349*10$^{-7}$&    \nodata& 8.069*10$^{-6}$&	 \nodata	 \\
          PN Pt 1& 2.048*10$^{-5}$& 3.606*10$^{-4}$& 5.539*10$^{-5}$& 8.610*10$^{-5}$&    \nodata&    \nodata&	 \nodata& 4.005*10$^{-6}$&    \nodata&    \nodata&	 \nodata	 \\

\hline
\end{tabular}
\vfill
\end{minipage}
\end{sidewaystable*}

\begin{sidewaystable*}
\begin{minipage}[t][180mm]{\textwidth}
Table 5 continued
\vspace{2cm}
\centering
\begin{tabular}{lccccccccccc}
	   \hline\hline
            Source&NII&OI&OII&OIII&NeIII&NeIV&SIII&ArV&ArIII&ArIV&NI\\
            \hline
	
	   PU Vul& 2.173*10$^{-5}$& 4.032*10$^{-5}$& 1.124*10$^{-5}$& 3.004*10$^{-4}$& 8.958*10$^{-5}$& 1.508*10$^{-4}$& 2.329*10$^{-6}$& 8.126*10$^{-7}$&3.322*10$^{-7}$& 3.522*10$^{-6}$& 8.831*10$^{-4}$\\
	   RR Tel& 4.075*10$^{-6}$& 6.376*10$^{-6}$& 1.343*10$^{-5}$& 2.577*10$^{-4}$& 3.697*10$^{-5}$& 3.280*10$^{-4}$& 1.723*10$^{-6}$& 6.723*10$^{-7}$& 4.959*10$^{-7}$& 2.975*10$^{-6}$&  \nodata    \\
	   RT Ser& 2.248*10$^{-5}$& 3.522*10$^{-4}$&  \nodata& 1.001*10$^{-4}$&	\nodata& 1.508*10$^{-4}$& 6.194*10$^{-6}$&    \nodata&    \nodata&	\nodata& 8.145*10$^{-4}$	      \\
	 SS73 122& 9.189*10$^{-6}$& 2.258*10$^{-5}$& 1.121*10$^{-5}$& 1.805*10$^{-4}$& 1.309*10$^{-5}$& 5.898*10$^{-6}$& 3.025*10$^{-6}$& 6.724*10$^{-7}$& 4.608*10$^{-7}$& 2.546*10$^{-6}$&  \nodata    \\
        SS73 129&    \nodata&    \nodata& 1.595*10$^{-4}$& 2.415*10$^{-4}$&    \nodata&    \nodata&    \nodata&	 \nodata&    \nodata&    \nodata&    \nodata \\
     SS73 141&    \nodata& 1.324*10$^{-4}$&   \nodata&	 \nodata&    \nodata&    \nodata&    \nodata& 3.145*10$^{-6}$&    \nodata&    \nodata&     \nodata \\
        SS73 29& 8.416*10$^{-7}$&    \nodata&   \nodata& 3.885*10$^{-5}$&    \nodata&    \nodata&     \nodata&	 \nodata& 1.465*10$^{-7}$&    \nodata&     \nodata \\
      SS73 71& 5.180*10$^{-6}$& 1.486*10$^{-5}$&    \nodata& 1.007*10$^{-4}$& 1.916*10$^{-5}$& 4.335*10$^{-5}$& 1.525*10$^{-6}$& 7.450*10$^{-7}$& 1.098*10$^{-8}$& 1.060*10$^{-5}$&   \nodata \\ 
      SS73 96& 1.013*10$^{-5}$& 1.129*10$^{-5}$& 2.122*10$^{-5}$& 1.506*10$^{-4}$& 5.674*10$^{-6}$& 1.923*10$^{-4}$& 3.247*10$^{-6}$& 1.551*10$^{-6}$& 1.802*10$^{-8}$& 8.404*10$^{-6}$&    \nodata \\
          Th 3-29&    \nodata& 4.277*10$^{-5}$& 1.271*10$^{-5}$& 1.504*10$^{-4}$&    \nodata&    \nodata&     \nodata&	 \nodata& 1.908*10$^{-7}$& 1.859*10$^{-6}$&      \nodata \\
       UU Ser& 5.009*10$^{-6}$&    \nodata&   \nodata& 3.773*10$^{-5}$&    \nodata&    \nodata&     \nodata&	 \nodata&    \nodata&    \nodata& 0.002409 \\
       V2416 Sgr& 7.463*10$^{-6}$& 1.576*10$^{-4}$& 9.246*10$^{-7}$& 8.516*10$^{-5}$& 5.861*10$^{-6}$&    \nodata& 5.966*10$^{-6}$&3.397*10$^{-6}$& 8.160*10$^{-9}$& 1.168*10$^{-7}$& 6.465*10$^{-5}$ \\
    V2506 Sgr&    \nodata&    \nodata&  \nodata&	 \nodata&    \nodata&    \nodata&    \nodata&	 \nodata&    \nodata&    \nodata&    \nodata \\
       V2601 Sgr& 1.678*10$^{-5}$& 1.770*10$^{-5}$& 1.230*10$^{-5}$& 5.759*10$^{-5}$& 1.169*10$^{-5}$& 2.135*10$^{-5}$& 4.774*10$^{-6}$& 4.048*10$^{-6}$&    \nodata&    \nodata&  \nodata \\
    V2756 Sgr& 2.773*10$^{-6}$& 6.385*10$^{-6}$& 2.903*10$^{-5}$& 1.609*10$^{-4}$&	 \nodata& 7.063*10$^{-5}$&       \nodata& 7.047*10$^{-7}$& 5.654*10$^{-7}$&    \nodata& 0.002434 \\
    V2905 Sgr&    \nodata&    \nodata& 7.786*10$^{-5}$& 3.812*10$^{-4}$& 3.743*10$^{-5}$&    \nodata&     \nodata&	 \nodata&    \nodata& 1.902*10$^{-4}$&     \nodata \\
        V343 Ser& 1.368*10$^{-6}$&    \nodata& 9.453*10$^{-7}$& 2.570*10$^{-5}$&    \nodata&    \nodata& 4.584*10$^{-8}$& 1.283*10$^{-6}$&	\nodata& 2.078*10$^{-8}$& 6.240*10$^{-6}$ \\
       V3804 Sgr& 4.426*10$^{-6}$& 2.158*10$^{-5}$& 1.026*10$^{-5}$& 1.198*10$^{-4}$& 2.214*10$^{-5}$&    \nodata & 2.639*10$^{-6}$& 1.426*10$^{-6}$& 4.156*10$^{-7}$& 1.139*10$^{-6}$&   \nodata \\
       V4018 Sgr& 1.399*10$^{-5}$&    \nodata& 9.158*10$^{-6}$& 1.022*10$^{-4}$& 9.888*10$^{-6}$& 1.013*10$^{-4}$&       \nodata& 1.998*10$^{-6}$&    \nodata&    \nodata& 0.002753 \\
       V4141& 4.566*10$^{-6}$& 1.154*10$^{-5}$& 3.239*10$^{-6}$& 2.173*10$^{-4}$& 1.744*10$^{-5}$& 2.562*10$^{-5}$& 1.422*10$^{-6}$& 4.299*10$^{-7}$& 3.017*10$^{-7}$& 1.828*10$^{-6}$& 0.001029 \\
    V919 Sgr& 9.766*10$^{-5}$&    \nodata& 4.403*10$^{-4}$& 3.969*10$^{-4}$& 8.841*10$^{-5}$& 5.536*10$^{-4}$&       \nodata&	 \nodata&    \nodata& 6.609*10$^{-4}$&  0.03935 \\
   WRAY 16 377& 5.147*10$^{-6}$&    \nodata& 3.001*10$^{-5}$& 1.593*10$^{-4}$& 1.230*10$^{-5}$&    \nodata&      \nodata&	 \nodata& 7.878*10$^{-8}$& 1.332*10$^{-4}$&      \nodata \\
        Y Cra&    \nodata& 9.477*10$^{-5}$&   \nodata& 2.525*10$^{-4}$& 3.029*10$^{-5}$& 2.976*10$^{-4}$& 5.461*10$^{-6}$&    \nodata& 9.731*10$^{-8}$& 1.408*10$^{-4}$&  \nodata \\
															       
\hline\hline
\end{tabular}
\vfill
\end{minipage}
\end{sidewaystable*}

%

   \begin{table*}
      \caption{Relative elemental abundances}
         \label{table5}
	 \centering
         \begin{tabular}{lllll||lllll}
            \hline\hline
            &Source &  N/O & Ne/O & Ar/O  &&Source &  N/O & Ne/O & Ar/O\\
            \hline
            1& K6-6& 0.801 & \nodata & 0.421         &   34&V4018  Sgr & 1.527& 0.097 &\nodata  \\  
	    4&Th 3-29&\nodata&\nodata&0.014			&   35&SS73  122 & 0.820 & 0.072&0.017 \\        
	    5&H 1-25 & \nodata &\nodata & 0.002      	&  38&Pt  1 & 0.369 & \nodata & \nodata \\
	    7&Hen  3-1410& 0.519&\nodata &\nodata    	  &	39&H  2-38 & 0.646 & 0.079 & 0.003 \\
	    8&AS  210 & 0.915 &0.074 & 0.003          & 40&V2756  Sgr & 0.095 & \nodata & \nodata \\ 
	    10&H  1-36 &0.724 &0.123 &0.011          &  42&HD  319167 & 0.416 & 0.156 & 0.016 \\
	    12&HK  Sco& 0.982 & 0.068 & \nodata      &	 44&V3804  Sgr & 0.431 & 0.185 & 0.013 \\
	    13&CL Sco & \nodata &0.161 & 0.025         & 45&Hen  3-1342 & \nodata & \nodata & 0.073 \\
	    15&Y CrA & \nodata & 0.119 & 0.558         & 48 & SS73 96 & 0.477 & 0.037 & 0.056 \\
	    16&Hen 2-171 & 2.443 & 0.101 & 0.009     	 &  49 & H 2-34 & 0.571 & \nodata & 0.035 \\
	    17&AE  Ara & 0.118 & 0.155 & \nodata     	  & 50&SS 73 71 & \nodata & 0.190 & 0.105 \\
	    18&V343 Ser & 1.447&\nodata&\nodata 	&	  51&CD 43 -14304 & 0.106 & 0.088 & \nodata \\ 
	    19&FN  Sgr  & 0.920 & 0.173 &\nodata     	  & 52& Hen 3-1761 & 0.416 & 0.136 &  \nodata \\ 
	    20&MWC  960 & 0.499 & \nodata & \nodata  	    & 53& RR Tel & 0.303 & 0.143 & 0.013 \\
	    22&MaC  1-9 & 0.890 &\nodata  &\nodata   	     &    55 & V919 Sgr & 0.222 & 0.222 & \nodata \\     
	    24&V2601  Sgr & 1.364 & 0.203 & \nodata  	     &        56 & Hen 3-863 & \nodata & 0.026& \nodata \\ 
	    26&ALS  2 & 0.945 & 0.963 &\nodata       	 &  58 & WRAY 16 377 & 0.171 & 0.077& 0.836 \\   
	    27&V4141  Sgr& 1.409 & 0.080 & 0.010     	  & 61 & AG Peg & 0.435 & 0.047 & \nodata \\
	    28&V2416  Sgr & \nodata& 0.069 & 0.001     	 &  62& AS 327 & 0.425 & \nodata & \nodata \\
	    30&Hen  3-1591& 3.329 &0.090  & 0.033    	 &  63 & FG Ser & 0.289 & 0.077 & \nodata \\
	    33&V2905  Sgr&  \nodata&0.098 &\nodata   	& 64 & PU Vul & 1.931 & 0.298 & 0.013 \\  
	    \\ 	  
	      
	       \hline
         \end{tabular}
 \end{table*}

  \begin{table}
      \caption{Mean Ne/O ratios for different samples}
         \begin{tabular}{lll}
            \hline\hline
            Sample &  Ne/O & Source\\
            \hline
          Bulge PNe  &  0.168 $\pm$ 0.070 & Escudero et al. (2004) \\  
          Disk PNe & 0.181 $\pm$ 0.129 & Maciel \& K\"oppen (1994) \\
          Disk and Bulge PNe & 0.223 $\pm$ 0.085 & Exter et al. (2004) \\
          This sample & 0.154 $\pm$ 0.034 & \\
            \hline
         \end{tabular}
 \end{table}

The mean values show that symbiotic stars have a similar Ne/O ratio to the bulge PNe from Escudero et al. (2004) or disk PNe form Maciel \& K\"oppen (1994). On the other hand, the mean Ne/O for the Exter et al. (2004) sample, which combines bulge and disk objects, is higher than the value from our sample. Within the dispersion, this is another indication that our sample contains both bulge and disk objects. 

As an additional test, we included in our analysis some symbiotics that are outside of our definition of bulge. The mean and standard deviation of Ne/O including these objects is 0.147 $\pm$ 0.027. Since we have only a few objects outside the bulge, the derived mean values are essentially the same.

As a following project we intend to observe the full sample of southern symbiotic stars to increase the statistics about physical parameters and abundances for this class of objects in different locations in the Galaxy.

 \begin{figure}
   \centering
   \includegraphics[width=8.8cm]{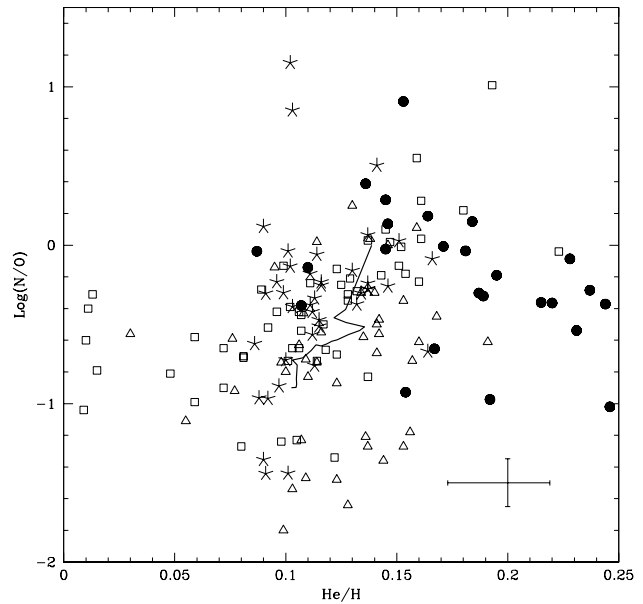}
    \caption{Log(N/O) vs. Log (He/H) for our sample(filled circles), planetary nebulae from Escudero \& Costa (2001) (triangles) and Escudero et al. (2004)(squares), Exter et al. (2004) (stars); solid line represents a model from Marigo (2001) with solar metallicity and mixing-length parameter $\alpha$ = 1.68 (see Marigo 2001 for details)}
   \end{figure}


\begin{acknowledgements}
      This work was supported by CNPq and FAPESP. G.J.M. Luna acknowledges CNPq for his graduate fellowship (Process 141805/2003-0). We acknowledge the comments and suggestions of the referee, which helped us to improve the final quality of this work.
\end{acknowledgements}


\begin{thebibliography}{}
\bibitem[Almog \& Netzer (1989)]{almog} Almog, Y., Netzer, H., 1989, MNRAS, 238, 57
\bibitem[Belckzy\`nski et al.(2000)]{belczynski} Belckzy\`nski, K., Mikolajewska, J., Munari, U., Ivison, R.J., Friedjung, M., 2000, A\&AS, 146, 407
\bibitem[Cardelli et al. (1989)]{cardelli} Cardelli, J., Clayton, G. and Mathis, J., 1989, ApJ, 345, 245
\bibitem[Clegg (1987)]{cleg} Clegg,  R. E. S., 1987, MNRAS, 229,31
\bibitem[Costa \& de  Freitas Pacheco 1994]{costa} Costa, R.D.D. \& de Freitas Pacheco, J.A., 1994, A\&A,285,998
\bibitem[Cuisinier et al.(1996)]{cuisinier} Cuisinier, F., Acker, A., K\"oppen, J., 1996, A\&A, 307, 215
\bibitem[Deuel \& Nussbaumer 1984]{deuel} Deuel, W. \& Nussbaumer, H., 1984, in: Rolfe E., Battrick, B.(eds) ESA SP-218, Proc. 4th Europeam IUE Conference,p.399
\bibitem[Escudero \& Costa 2001]{escudero} Escudero, A.V. \& Costa, R.D.D., 2001,A\&A, 380, 300
\bibitem[Escudero et al. 2004]{escudero1} Escudero, A.V., Costa, R.D.D., Maciel, W.J., 2004, A\&A, 414, 211
\bibitem[Exter et al. 2004]{exter} Exter, K.M., Barlow, M.J., Walton, N.A., 2004, MNRAS, 349, 1291
\bibitem[de Freitas Pacheco \& Costa (1992)]{pacheco}de Freitas Pacheco, J.A., Costa, R.D.D., 1992, A\&A, 257, 619
\bibitem[Gutierrez Moreno et al. (1992)]{moreno2} Gutierrez-Moreno, A., Moreno, H.\& Feibelman, W. A.,1992,ApJ, 395,295
\bibitem[Gutierrez Moreno \& Moreno 1996]{moreno1} Gutierrez Moreno \& Moreno,1996, PASP, 108, 972
\bibitem[Gutierrez Moreno \& Moreno 1999]{moreno} Gutierrez Moreno \& Moreno,1999, PASP, 111, 571
\bibitem[Kingdon \& Ferland 1995]{kingdon} Kingdon, J. \& Ferland, G.J., 1995, ApJ, 442, 714
\bibitem[Maciel \& K\"oppen 1994]{maciel} Maciel, W.J., K\"oppen,J., 1994, A\&A, 282, 436 
\bibitem[Marigo 2001]{marigo} Marigo, P., 2001, A\&A, 370,194
\bibitem[Medina Tanco \& Steiner 1995]{medina} Medina Tanco, G.A. \& Steiner, J.E., 1995, AJ,109,1770 
\bibitem[Michalitsianoset al. (1982)]{micha} Michalitsianos, A. G.; Feibelman, W. A.; Hobbs, R. W.; Kafatos, M., 1982, ApJ, 253, 735
\bibitem[Mikolajewska et al. 1997]{mikola} Mikolajewska, J., Acker, A., Stenholm, B., 1997, A\&A, 327, 191
\bibitem[Munari \& Buson (1993)]{munari93} Munari, U. \& Buson, L. M., 1993, MNRAS, 263, 267
\bibitem[Munari \& Buson (1994)]{munari93} Munari, U. \& Buson, L. M., 1994, A\&A, 287, 87
\bibitem[Netzer (1975)]{netzer}Netzer, H., 1975, MNRAS, 171, 395
\bibitem[Nussbaumer et al. 1988]{nussbaumer} Nussbaumer, H, Schmid, H M., Vogel, M, Schild, H., 1988, A\&A, 198, 179
\bibitem[Pereira 1995]{pereira95} Pereira, C.B., 1995, \textit{Astron. Astrophys. Suppl. Ser.},11, 471
\bibitem[Pereira 1998]{pereira98} Pereira, C. B., Landaberry, S. J. C. \& Junqueira, S., 1998, A\&A, 333, 658
\bibitem[Pereira et al.(2002)]{pereira} Pereira, C.B., Franco, C.S., de Ara\'ujo, F.X., 2002, A\&A, 385, 900
\bibitem[Osterbrock 1989]{osterbrok} Osterbrock, D.E., Astrophysics of Gaseous Nebulae and Active Galactic Nuclei (University Science Books, Mill Valley, CA, 1989).
\bibitem[Schmid, H. M. \& Nussbaumer, H. (1993)]{nuss93} Schmid, H. M.\& Nussbaumer, H., 1993, A\&A,268,159
\bibitem[Schmid \& Schild 1990]{schmidt} Schmid, H.M., Schild, H., 1990, MNRAS, 246, 84
\bibitem[Shaw \& Dufour (1994)]{shaw} Shaw, R. A. \& Dufour, R. J., 1994, ASP Conf. Ser., 61,327
\bibitem[Simon 2003]{simon} Simon, V., 2003, A\&A, 406, 613
\bibitem[Stauffer 1984]{stauffer} Stauffer, J.R., 1984, ApJ, 280, 695
\bibitem[Willson et al. 1984]{willson} Willson, L.A., Wallerstein, G., Brugel, E.W., Stencel, R.E., 1984, A\&A, 133, 154  

\end{thebibliography}
\end{document}